\documentclass[12pt]{iopart}
\usepackage{epsfig}
\newcommand{\ben}{\begin{eqnarray}}
\newcommand{\een}{\end{eqnarray}}

\begin{document}
\title{Transverse momentum broadening of vector bosons 
in heavy ion collisions at the LHC}
\author{Zhong-Bo Kang$^1$ and Jian-Wei Qiu$^{1,2}$}
	
\address{$^1$Department of Physics and Astronomy, 
                    Iowa State University, 
                    Ames IA 50011, USA\\
	      $^2$Physics Department, Brookhaven National Laboratory, 
		  Upton, NY 11793, USA}
\ead{kangzb@iastate.edu {\rm and} jwq@iastate.edu}    

\begin{abstract}
We calculate in perturbative QCD the transverse momentum broadening 
of vector bosons in heavy ion collisions at the Large Hadron 
Collider (LHC).  We predict transverse momentum broadening of 
$W/Z$ bosons constructed from their leptonic decay channels, 
which should be a clean probe of initial-state medium effect.  
We also predict the upper limit of transverse momentum broadening 
of J/$\psi$ and $\Upsilon$ production as a function of N$_{\rm part}$
at the LHC energy.
\end{abstract}


Nuclear transverse momentum broadening of heavy vector bosons
($\gamma^*$, $W/Z$, and heavy quarkonia) is defined 
as a difference between the averaged transverse momentum square
measured in nuclear collisions and that measured in collisions of 
free nucleons,
\ben
\Delta \langle q_T^2\rangle_{AB} 
\equiv 
\langle q_T^2\rangle_{AB}-
\langle q_T^2\rangle_{NN}
\approx 
\int dq_T^2\, q_T^2\, \frac{d\sigma^{(D)}_{AB}}{dq_T^2}
\left/
\int dq_T^2\, \frac{d\sigma_{NN}}{dq_T^2}
\right. \, .
\label{dqt2}
\een
Since single scattering is localized in space, the broadening is
a result of multiple parton scattering, and is a good
probe for nuclear medium properties.  Because the mass scale of 
the vector bosons is much larger than the characteristic momentum
scale of the hot medium, the broadening is likely dominated 
by double partonic scattering as indicated in Eq.~(\ref{dqt2}).  
The broadening caused by the double scattering can be 
systematically calculated in terms of high twist formalism in 
QCD factorization \cite{Luo:1993ui,Guo:1998rd}. 

\begin{figure}[hbt]
\centering
\psfig{file=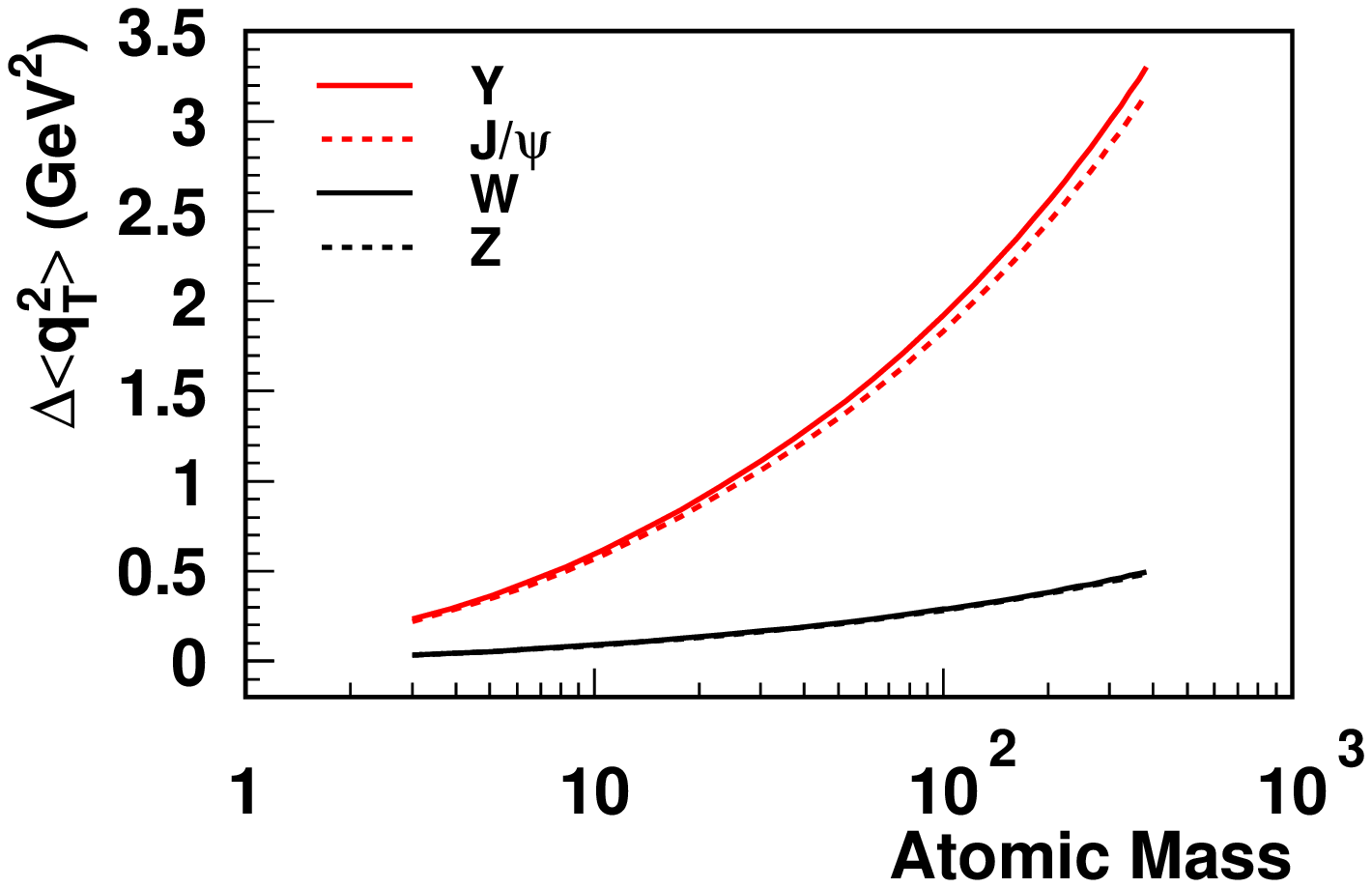,height=2in}
\psfig{file=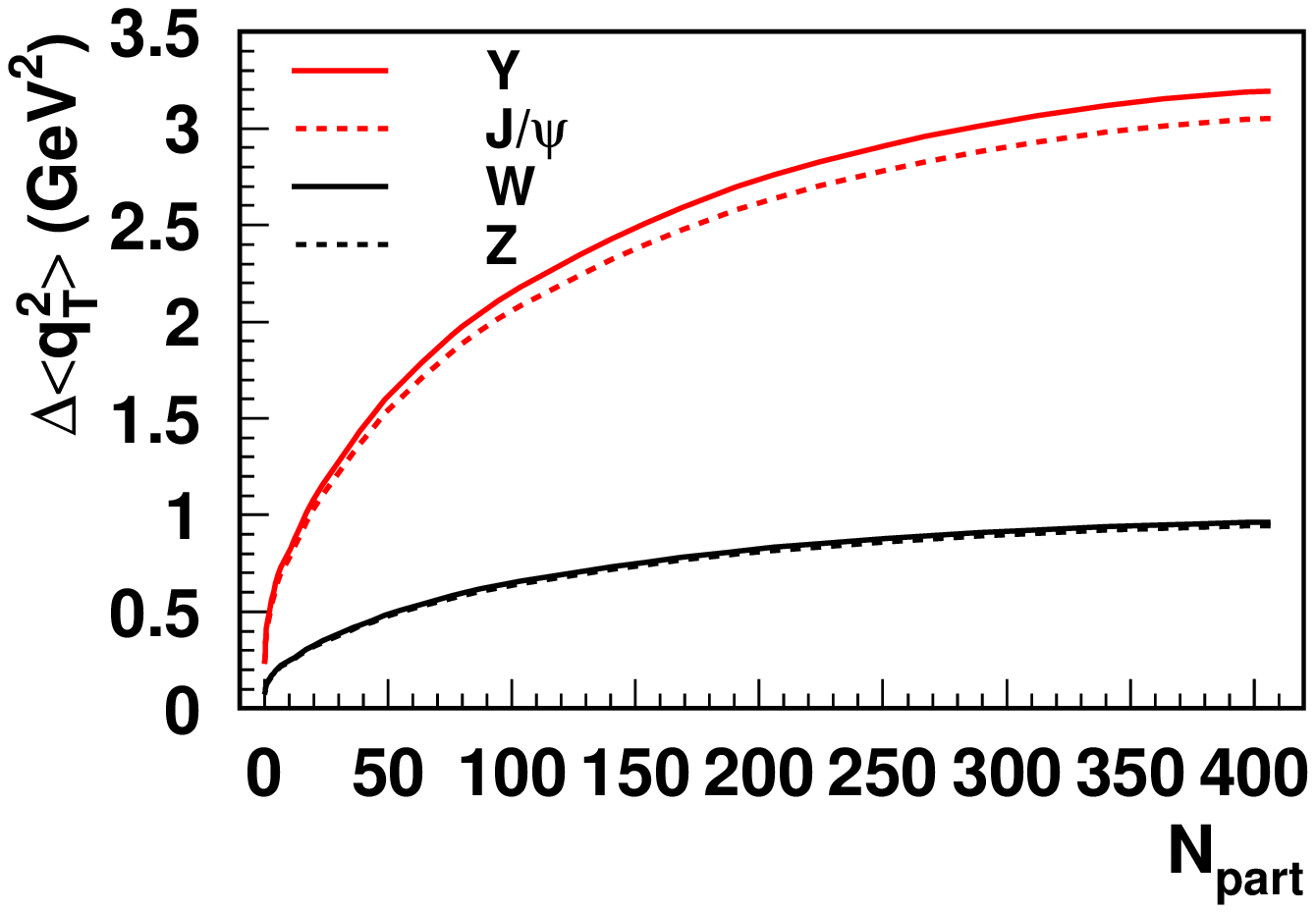,height=2in}
\caption{Predicted broadening (maximum broadening) for 
         $W$ and $Z$ (J/$\psi$ and $\Upsilon$) production 
         in p-A (left) and Pb-Pb (right) collisions 
         at $\sqrt{s}=5500$ GeV.}
\label{vbbroaden}
\end{figure}

At the LHC energies, a lot $W$ and $Z$, and 
J/$\psi$ and $\Upsilon$ will be produced.  Most reconstructed 
$W/Z$ bosons will come from their leptonic decays.  
Their transverse momentum broadening is a result of purely 
initial-state multiple scattering.  By calculating the double 
scattering effect, we obtain \cite{Guo:1998rd,Kang:2007xx}
\ben
\Delta\langle q_T^2\rangle_{pA}^W
=\frac{4\pi^2\alpha_s(M_W)}{3} \lambda^2_W A^{1/3} \, ,
\quad
\Delta\langle q_T^2\rangle_{pA}^Z
=\frac{4\pi^2\alpha_s(M_Z)}{3} \lambda^2_Z A^{1/3}
\label{broaden}
\een
for hadron-nucleus collisions.  The $\lambda^2 A^{1/3}$ 
in Eq.~(\ref{broaden}) was introduced 
in \cite{Luo:1993ui} as a ratio of nuclear four parton correlation 
function over normal parton distribution.  The $\lambda$ 
is proportional to the virtuality or transverse momentum
of soft gluons participating in the coherent double scattering.  
For collisions with a large momentum transfer, $Q$, 
the $\lambda^2$ should be proportional to $\ln(Q^2)$ 
\cite{Kang:2007xx} and 
the saturation scale $Q_s^2$ if the active parton $x$ is small.
By fitting Fermilab E772 Drell-Yan data, it was found that 
$\lambda^2_{\rm DY}\approx 0.01$GeV$^2$ at $\sqrt{s}=38.8$GeV 
\cite{Guo:1998rd}.  From the $\lambda^2_{\rm DY}$, we
estimate the value of $\lambda^2$ for production of a vector boson 
of mass $M_V$ at the LHC energy as 
\ben
\lambda^2_V({\rm LHC})\approx
\lambda^2_{\rm DY}\
\frac{\ln(M_V^2)}{\ln(Q_{\rm DY}^2)}\,
\left(\frac{M_V/5500}{Q_{\rm DY}/38.8}\right)^{-0.3}\, ,
\label{lambda}
\een
where we used $Q_s^2 \propto 1/x^\delta$ with $\delta\approx 0.3$ 
\cite{Golec-Biernat:1998js} and 
$\sqrt{s}=5500$~GeV for the LHC heavy ion 
collisions.  For an averaged $Q_{\rm DY}\sim 6$~GeV, we obtain
$\lambda^2_{W/Z}\approx 0.05$ at the LHC energy.  
We can also apply our formula in Eq.~(\ref{broaden}) 
to the broadening in nucleus-nucleus collisions by replacing 
$A^{1/3}$ by an effective medium length $L_{eff}$. 
We calculate $L_{eff}$ in Glauber model with inelastic 
nucleon-nucleon cross section $\sigma_{NN}^{in}=70$mb at the 
LHC energy.  We plot our predictions (lower set curves) 
for the broadening of $W/Z$ bosons in Fig.\ref{vbbroaden}.  

Heavy quark pairs are produced at a distance scale much less than
the physical size of heavy quarkonia in high energy 
collisions.  The pairs produced in heavy ion collisions
can have final-state interactions before bound quarkonia could
be formed. We found \cite{Kang:2007xx} that with both initial- and 
final-state double scattering, the broadening of heavy quarkonia 
is close to $2C_A/C_F$ times the Drell-Yan broadening
in proton-nucleus collision, which is consistent with existing data
\cite{Peng:1999gx}.  If all soft gluons of heavy ion beams are stopped to 
form the hot dense medium in nucleus-nucleus collisions, final-state
interaction between the almost stationary medium and the fast moving 
heavy quarks (or quarkonia) of transverse momentum $q_T$ is 
unlikely to broaden the $q_T$ spectrum, instead, it is likely to 
slow down the heavy quarks (or quarkonia) \cite{Kang:2007xx}.  
From Eq.~(\ref{lambda}). we obtain
$\lambda^2_{{\rm J/}\psi}\approx 0.035$, and 
$\lambda^2_{\Upsilon}\approx 0.049$ at the LHC energy; and we 
predict the maximum broadening for J/$\psi$ and $\Upsilon$ production
(upper set curves) in Fig.~\ref{vbbroaden}. 

This work is supported in part by the US Department of Energy  
under Grant No. DE-FG02-87ER40371 and contract number 
DE-AC02-98CH10886.


\end{document}